\documentclass[sigconf]{acmart}

\AtBeginDocument{%
  \providecommand\BibTeX{{%
    \normalfont B\kern-0.5em{\scshape i\kern-0.25em b}\kern-0.8em\TeX}}}

\copyrightyear{2022}
\acmYear{2022}
\setcopyright{rightsretained}
\acmConference[UIST '22]{The 35th Annual ACM Symposium on User Interface Software and Technology}{October 29-November 2, 2022}{Bend, OR, USA}
\acmBooktitle{The 35th Annual ACM Symposium on User Interface Software and Technology (UIST '22), October 29-November 2, 2022, Bend, OR, USA}
\acmDOI{10.1145/3526113.3545660}
\acmISBN{978-1-4503-9320-1/22/10}

\usepackage{color}
\usepackage{textcomp}
\usepackage{multirow}
\usepackage{subcaption}
\usepackage{bm}
\usepackage{wrapfig}

\newcommand{\xhdr}[1]{\vspace{1mm}\noindent{{\bf #1.}}}

\newcommand{\new}[1]{\textcolor{black}{#1}}

\newcommand{\system}{\textsc{Threddy }}
\newcommand{\threddy}{\textsc{Threddy}}

\newcommand{\baseline}{\textsc{Google Docs }}

\newcommand{\bs}{\textsc{Google Docs}}

\newcommand*\cirnum[1]{\raisebox{.5pt}{\textcircled{\raisebox{-.9pt} {#1}}}}

\newcommand{\ul}[1]{\begin{itemize}#1\end{itemize}}
\newcommand{\li}[1]{\item{#1}}
\DeclareUnicodeCharacter{202F}{\,}

\begin{document}

%%
%% The "title" command has an optional parameter,
%% allowing the author to define a "short title" to be used in page headers.
\title[\threddy]{\threddy: An Interactive System for Personalized Thread-based Exploration and Organization of Scientific Literature}

\author{Hyeonsu B. Kang}
% \authornote{Both authors contributed equally to this research.}
\email{hyeonsuk@cs.cmu.edu}
\orcid{0000-0002-1990-2050}
\affiliation{%
 \institution{Carnegie Mellon University}
 \streetaddress{5000 Forbes Ave}
 \city{Pittsburgh}
 \state{Pennsylvania}
 \postcode{15213}
 \country{USA}
}

\author{Joseph Chee Chang}
% \authornote{Both authors contributed equally to this research.}
\email{josephc@allenai.org}
\orcid{0000-0002-0798-4351}
\affiliation{%
  \institution{Allen Institute for AI}
  \streetaddress{2157 N Northlake Way \#110}
  \city{Seattle}
  \state{WA}
  \postcode{98103}
  \country{USA}
}

\author{Yongsung Kim}
% \authornote{Both authors contributed equally to this research.}
\email{yongsung@cmu.edu}
\orcid{0000-0002-2026-8050}
\affiliation{%
 \institution{Carnegie Mellon University}
 \streetaddress{5000 Forbes Ave}
 \city{Pittsburgh}
 \state{Pennsylvania}
 \postcode{15213}
 \country{USA}
}

\author{Aniket Kittur}
% \authornote{Both authors contributed equally to this research.}
\email{nkittur@cs.cmu.edu}
\orcid{0000-0003-4192-9302}
\affiliation{%
 \institution{Carnegie Mellon University}
 \streetaddress{5000 Forbes Ave}
 \city{Pittsburgh}
 \state{Pennsylvania}
 \postcode{15213}
 \country{USA}
}

%%
%% By default, the full list of authors will be used in the page
%% headers. Often, this list is too long, and will overlap
%% other information printed in the page headers. This command allows
%% the author to define a more concise list
%% of authors' names for this purpose.\renewcommand{\shortauthors}{Trovato and Tobin, et al.}

% \author{Leave Authors Anonymous for Submission}

%%
%% The abstract is a short summary of the work to be presented in the
%% article.
\begin{abstract}
% Reviewing the literature to understand relevant threads of past work is a critical part of research and vehicle for learning. However, as the scientific literature grows the challenges for users to find and make sense of the many different threads of research grow as well. Previous work has helped scholars to find and group papers with citation information or textual similarity using standalone tools or overview visualizations. Instead, in this work we explore a tool integrated into users' reading process that helps them with leveraging authors' existing summarization of threads, typically in introduction or related work sections, in order to situate their own work's contributions. To explore this we developed a prototype that supports efficient extraction and organization of threads along with supporting evidence as scientists read research articles. The system then recommends further relevant articles based on user-created threads. We evaluate the system in a lab study and find that it helps scientists to create and relate threads, collect relevant papers and clips, and discover interesting new articles to further grow threads.
Reviewing the literature to understand relevant threads of past work is a critical part of research and vehicle for learning. However, as the scientific literature grows the challenges for users to find and make sense of the many different threads of research grow as well. Previous work has helped scholars to find and group papers with citation information or textual similarity using standalone tools or overview visualizations. Instead, in this work we explore a tool integrated into users' reading process that helps them with leveraging authors' existing summarization of threads, typically in introduction or related work sections, in order to situate their own work's contributions. To explore this we developed a prototype that supports efficient extraction and organization of threads along with supporting evidence as scientists read research articles. The system then recommends further relevant articles based on user-created threads. We evaluate the system in a lab study and find that it helps scientists to follow and curate research threads without breaking out of their flow of reading, collect relevant papers and clips, and discover interesting new articles to further grow threads.
\end{abstract}
%%
%% This command processes the author and affiliation and title
%% information and builds the first part of the formatted document.
\maketitle

%%
%% The code below is generated by the tool at http://dl.acm.org/ccs.cfm.
%% Please copy and paste the code instead of the example below.
%%


%% Keywords. The author(s) should pick words that accurately describe
%% the work being presented. Separate the keywords with commas.

\section{Introduction}
\begin{figure*}[t]
    \includegraphics[width=.95\textwidth]{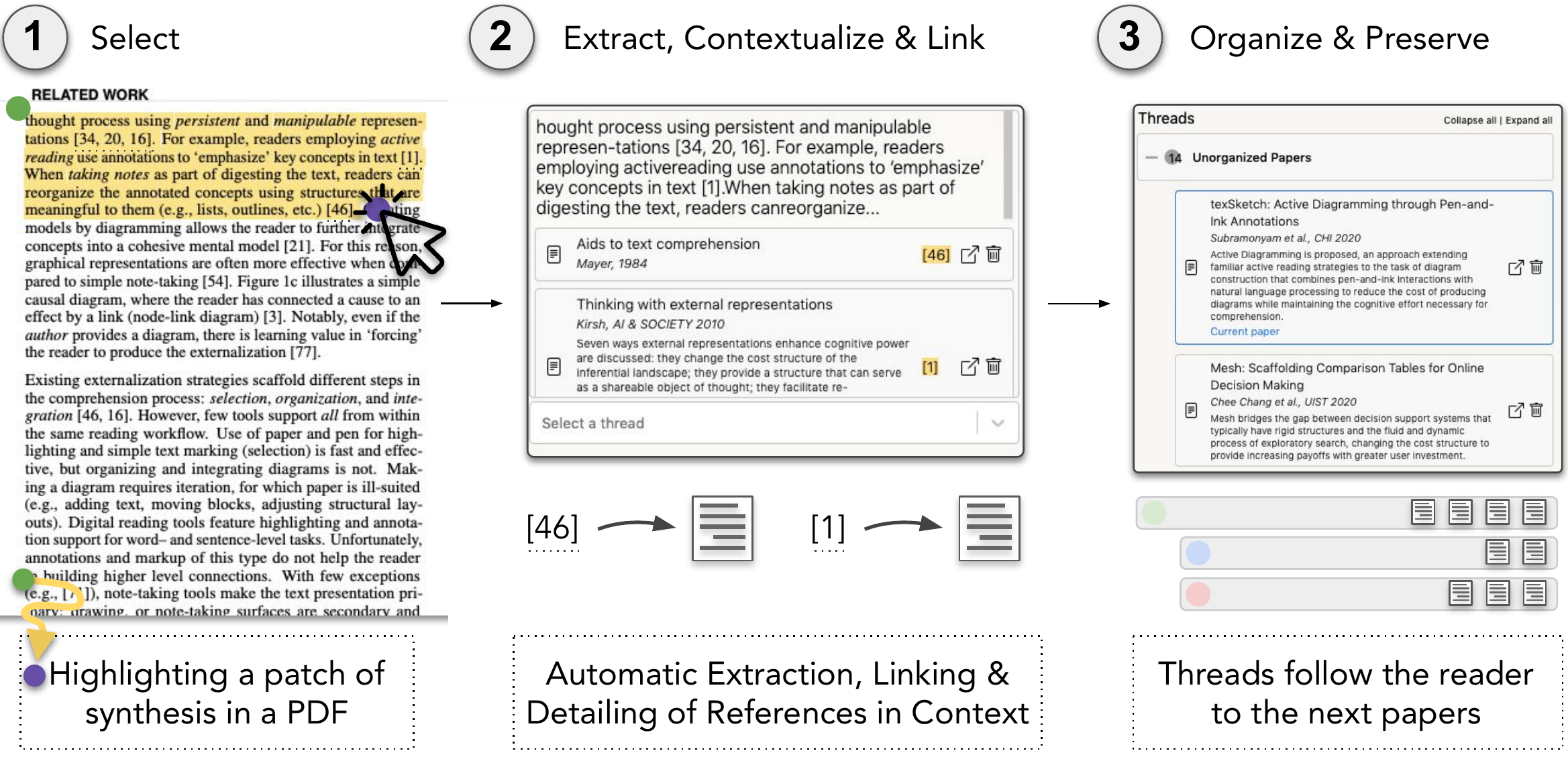}
    \vspace{-1em}
    \caption{Thread creation and organization while reading on \threddy: \cirnum{1} The reader highlights a useful patch of text that interleaves references in a citation context. \cirnum{2} The system extracts the referenced papers from the highlighted text, link them to the citation context, and show the resulting data to the reader. \cirnum{3} The reader creates a new thread together with the citation context and extracted references. \new{Alternatively,} the user can add the context \new{only as a relevant} clip to an existing thread, or add  the extracted references to it. The threads and their context follow the reader as she \new{continues on reading} other papers.}
    \vspace{-1em}
    \label{fig:main}
\end{figure*}

Reviewing the literature to understand relevant threads of research is a critical part of scientific research and serves as a research facilitator and a vehicle for learning~\cite{bruce1994research}. For example, a scholar trying to understand the history of tools that support scientific literature review might learn about research threads including overview visualizations based on citation networks; augmentative interfaces for active reading; collection tools that help scholars organize their papers; and so forth. Understanding prior threads of research is critical to building on past work, finding inspirations for new innovation, and positioning contributions in the appropriate research context.

However, as the scientific literature grows the challenges for users to find and make sense of the many different threads grow as well. Finding and keeping track of papers within a single thread can be challenging, requiring users to traverse references and citations, read through introductions and related work sections, and search across various keywords to avoid missing important work. Exacerbating the problem, scholars are often interested in multiple threads that are relevant to their work, with each often branching into multiple sub-threads as the example described above of literature review support tools demonstrates.

An effective strategy for `shortcutting' the cumbersome process of assembling research threads is to harvest and build on the work that other scholars have already done in assembling them. This process, commonly used among scholars~\cite{bruce1994research,CiteSense}, involves reading through papers (typically in the introductions and related work sections) to find how the authors have compressed and summarized the threads of research relevant to their papers in order to situate their own work's contributions. These predigested threads provide scaffolding for users in assembling their own, both in terms of the references cited as well as the citing text describing those references. By following the most relevant references, finding more citing texts, and further chaining through papers, scholars can more quickly assemble an overview of the research threads in an area than by searching and collecting individual papers alone.

However, even this process of inferring threads is cumbersome. Consider the following scenario in which a scientist is learning from a new research paper. Quickly skimming the introduction, she may identify a useful patch of text in it which describes a research thread she would like to explore further. This patch of text (e.g., a sentence, a paragraph, or a section) often contains a number of citations pointing to related work and describes their relation which provides a helpful context. Deciding to save this context and follow up on one of the references requires her to first jump between the references section and the citation context in \new{the} body text to link the reference notation she wants to follow up with \new{to} the actual title and URL of the paper. Next she needs to locate the actual content of the paper, perhaps by querying its title on a search engine. Finally, she may make notes of the found paper and save it for future reference. She needs to repeat this multiple times for each patch of text she finds interesting, the cost of which compounds quickly. The reference notation used in papers may also differ, sometimes providing little context about what they are (e.g., numbers in a bracket \new{such as `[1]'}), and this lays an even more \new{burdensome} task of \new{correctly} mapping and linking papers, which (using one of our study participant's words) can be ``\textit{a real, damaging context break.}'' In addition, she might after all end up finding that the actual content of the cited paper to be \new{irrelevant or} uninteresting, in which case she must resume her \new{flow of} reading by re-building \new{the lost} context and \new{previous} threads of thoughts.

As she moves to other papers and collecting more patches of relevant information, it may become clear that they relate to each other along some threads she created, and this needs to be captured. In order to achieve this, she first needs to look through her own notes, threads, and references that she may have loosely organized, and this quickly becomes a sizable sub-task that once again requires her to break out of her flow of reading in order to complete it. Furthermore, it is easy to forget which references are already looked at and which are new that need to be processed, which may incur additional friction to the process. She may also have multiple threads she would like to follow up on at any given time, without having an easy way of keeping track of them while she is reading a paper. Using multiple tabs or groups of them might be an intuitive way to organize articles into related threads (albeit its potential for creating a tab overload~\cite{Chang2021WhenTT,chang2021tabs,greis2017increasing}), but this does not help with maintaining patches of citation context that described the threads. Finally, once relevant patches of text are collected for each thread from source papers, there is no easy way to use this information to find additional relevant papers that she may use to further grow the threads.

\new{While significant research has focused on supporting scientific literature search and collection, there is relatively little support for users building threads during reading.} For example, \textit{Papers101}~\cite{choe2021papers101} aims at supporting early-stage scholars' discovery of literature by recommending relevant but unused keywords for query. \new{Alternatively,} \textit{Apolo}~\cite{apolo} \new{adopts and applies} the Belief Propagation~\cite{bp} \new{algorithm} on the citation network \new{in a novel way} to support progressive retrieval of papers given a set of papers that the user has collected thus far. Once a list of recommendations are curated, \new{systems such as} \textit{PaperQuest}~\cite{paperquest} provide support for triaging what to read next, and \textit{VisualBib}~\cite{dattolo2021assisting} aims at providing a more holistic support for managing the growing user-curated bibliographies. However, none of these systems provides a mechanism for leveraging the data scholars have collected and assembled into threads \new{while reading research articles} to recommend further relevant papers to continue growing the threads.

In this paper we aim to address this gap in the literature by developing \threddy, a system that supports users with collecting patches of text \new{in research articles} that contain pre-digested syntheses by other authors (i.e., \new{a} useful citation context along with automatically extracted references), and helps them assemble personal \new{research} threads using clippings of others' pre-digested threads. Different from \new{the related prior} work, \system does not create a new information environment nor an application context that requires users to context switch away from their natural flow of reading, but instead seamlessly integrates the support it provides into the user's in-situ context of reading. We evaluate \system in a controlled lab study with 9 scientists conducting literature review in \new{personalized} domains, and demonstrate \new{how it increases} \new{users'} effectiveness \new{in} leveraging pre-digested syntheses by other authors to enrich their own threads, \new{decreases} the cost of frequent context switching they would have experienced in a similar task without the tool, \new{and heightens users'} flow state while conducting a literature review.

\section{Related Work}
\new{Scholars exploring and reviewing the relevant literature are challenged by} the immense scale and dynamic changes in the literature~\cite{jinha2010article,bornmann2015growth,noorden_doubling_output,tenopir2009electronic,rowland2002overcoming,megwalu2015academic}, \new{which could overload them with information~\cite{miller1960information}, making} the effective allocation of attention~\cite{simon1996designing} an imperative. The challenge may be especially pronounced for \new{scholars tasked with} finding relevant papers, organizing them into multiple, evolving research threads, and growing the threads with recent papers from the literature. \new{It is therefore unsurprising that} scholars' \new{discovery} experience with the literature is often characterized as tedious, scattered, and relied upon chance~\cite{landhuis2016scientific,breitinger2019too}. \new{To mitigate this, several systems were developed to help scholars making sense of research articles which we organize into the research threads below.}

\subsection{\new{Tools that Support Interaction within and across Documents}}
\new{\textit{Passages}~\cite{passages} proposed an approach to `reify'~\cite{generative_theories_of_interaction} ephemeral user text selections into persistent objects that can be shared across multiple applications and in various representations such as a canvas or a spreadsheet view. In an evaluation with scientists using a set of six papers, \textit{Passages} was favored over the baseline and shown to encourage reuse of the text and structure collected during the document interaction. In comparison, \textit{texSketch}~\cite{texsketch} developed an interactive canvas for deeply engaging with individual documents by supporting concept diagramming while reading. Other work in code~\cite{omnicode} and document programming~\cite{iLatex} also showed how dual representations dynamically generated from user selected portions of content could improve end-user comprehension, engagement, and learning. Furthermore, Wang et al. examined how visualization of the narrative structure within the related work sections of a paper could improve comprehension and curation~\cite{wang2016guided}. Relatedly, \textit{``Metro Maps of Science''}~\cite{shahaf2012metro} used the metro map metaphor to describe the narrative structure of scientific progress. For a collection of documents, \textit{IntentStreams} proposed an approach to group and visualize documents relevant to a search query into streams~\cite{andolina2015intentstreams}, while \textit{Apolo}~\cite{apolo} and \textit{IdeaHound}~\cite{ideahound} used a 2-D spatial arrangement to quickly overview the clusters of similar documents.}

\subsection{\new{Tools that Augment Reading Experiences}}
Reading an individual paper may already involve a high cognitive cost such as unfamiliar terms and `nonce' words~\cite{head2021augmenting}, complex formulae~\cite{head2022math} and barriers to domain-specific knowledge~\cite{hillesund2010digital,bazerman1985physicists,nicholas2010researchers}. Therefore, prior work has explored ways to mitigate this cost. One thread of research focused on cross-referencing within \new{a single document} to improve readability (e.g.,~\cite{rachatasumrit2022citeread,Scim,PaperPlain}). \new{A sub-thread of this is helping users more efficiently navigate to and from different parts of a table and the corresponding snippets of text~\cite{Kong2014ExtractingRB,Badam2019ElasticDC,Kim2018FacilitatingDR}. In comparison, \textit{NB}~\cite{NB} binds contextually relevant discussions to the margins of a document students are reading in a class to enhance their engagement and learning.} \new{These systems and their findings show the importance of reducing the cognitive cost that may be associated with reading complex text such as research papers. At the same time, it alludes to the potential benefit of augmenting reading interfaces in a way that readers can re-use and build upon other authors' pre-digested synthesis from multiple papers in a contextually relevant way, without breaking out of their flow of reading, to better retain focus and their mental models of the structure of relevant research threads.}

\subsection{\new{Tools for the Literature Review Workflow}}
\new{Several work also modeled the literature review process as an iterative process of curation and querying of papers, and developed systems that can support this process in a standalone information environment.} In \textit{PaperQuest}~\cite{paperquest} users could input query papers to receive other relevant papers based on citation relationships. \textit{Sturm}~\cite{sturm2019design} surveyed requirements of systematic literature search systems, and designed the \textit{LitSonar} system which provides support for querying over multiple sources of document streams using nested queries. \textit{LitSense}~\cite{litsense} \new{developed} multiple visualizations \new{based on} citation relations, and enabled filtering and querying operations for homing in on specific references that the end-users may further explore. \textit{CiteSense}~\cite{CiteSense} developed an information-rich environment which provides various features for searching, appraising, and managing the different tasks involved in a literature review.
\new{While prior work in supporting scholarly literature review primarily focused on a standalone information environment detached from users' flow of reading, recent notable exceptions in online sensemaking support tools are} \textit{ForSense}~\cite{rachatasumrit2021forsense} \new{and \textit{Fuse}~\cite{kuznetsov_fuse_2022}. Both systems included a sidebar view in the margins of the webpage the user is reading, and allowed} users to organize clips and notes into clusters \cite{rachatasumrit2021forsense} or fluid and hierarchical structures \cite{kuznetsov_fuse_2022}.

\section{Usage Scenario and System Design}
\begin{figure*}[t]
    \centering
    \includegraphics[width=.95\textwidth]{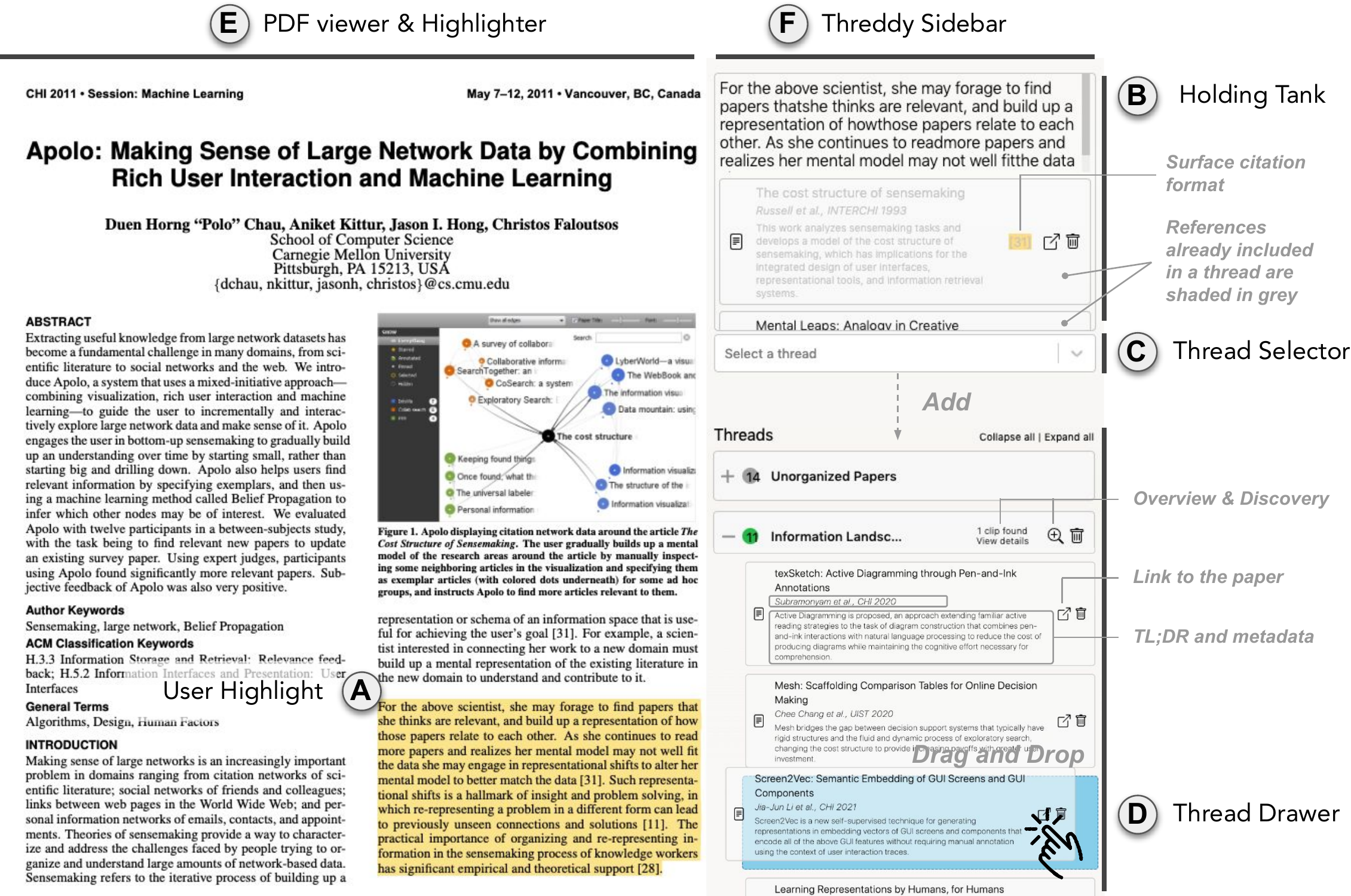}
    \vspace{-1em}
    \caption{The design of \system consists of two primary panels, (E) PDF viewer \& Highlighter and (F) Sidebar. When the user (A) highlights text in the PDF, its content and references are found and temporarily stored in the (B) holding tank. The user can review the content of the holding tank and clean up any errors in the automated extraction and linking of references. When the content looks good, the user can either type in the (C) thread selector to create a new thread, or choose an existing thread. Choosing an option selectively activates buttons for 1) creating a new thread with the references, 2) only adding the extracted references to the chosen thread, and 3) only adding the content of the holding tank as a clip to the chosen thread. Once the user chooses the intended operation, the (D) thread drawer's content is updated to reflect the change. The user can interact with the thread drawer to organize and re-organize its content. The changes are stored and persist across other paper PDFs.}
    \vspace{-1em}
    \label{fig:main}
\end{figure*}

\xhdr{\new{Usage Scenario}} \new{We first illustrate how an end-user, Sam, would interact with \system to conduct a literature review. Sam is reading a paper when she encounters an interesting patch of text (see also figure~\ref{fig:main}). The two paragraphs in the related work section describe particularly relevant research threads that she wants to follow up on and save the references included in them for a deeper look. She quickly highlights the paragraph,} which triggers \system to search for references included in it, \new{automatically extract the metadata corresponding to each, and present them as interactive objects in the sidebar. Sam glances over the linked references and removes a couple of them which seemed less relevant based on their titles and TL;DR summaries, while keeping the rest. The extracted references and the context seem to form a good grouping for revisiting later, so she adds them as a new thread labeled `\textit{Reifying ephemeral user interaction (e.g., text selections)}'. She continues reading the paragraphs and repeatedly extracts and saves additional context, references, and threads such as `\textit{Systems for constructing narrative structures in sciences}', and `\textit{Systems that augment document margins}'. While these new groupings are helpful, she is not yet confident if the newly added threads and the hierarchy would make sense in light of additional references she would find later. She is also a little worried that the saved references may not provide a good coverage on the topic.}

\new{This leads her to click on one of the newly saved thread for a detailed look. In it, she finds a panel that shows additional references which cite several of the curated papers for the thread which seem relevant and useful. In particular, she finds three recent papers that cited a few of the papers she saved for the thread at the top of the panel. Seeing these papers leads her to create another thread with them. This also leads to a change in her mental model around the higher level research thread; she realizes there is a parent-level concept that aptly contains two of the threads she curated until now as its children. She creates it and nests the two threads as its children to better capture her updated mental model. She then collapses the parent thread to declutter the view and focus her attention to the other thread -- `\textit{Systems that augment document margins}' -- that she has not yet looked at in detail. She finds one of the papers included in the thread particularly interesting, and clicks on it to switch to the PDF. Though her attention shifts towards reading the new paper, she nevertheless maintains her awareness of the current research thread she is interested in from the persisting content of the sidebar view on the right of the new PDF. At the top of the sidebar shows the most recent thread she made changes to, which shows the hierarchy of threads she has been organizing thus far. In the new paper's related work section, she finds interesting new examples of systems from the prior work and the corresponding threads describing how document margins might be augmented in different ways to improve learner discussion, engagement, or comprehension of technical documents. She adds these papers to `\textit{Systems that augment...}' with the context and sub-threads corresponding to the specific ideas described in the paper.}

\xhdr{\new{System Design Rationale}} \new{Literature review is a complex task that most likely spans long durations, may be interrupted by other tasks~\cite{searchbar}, and is often initiated and resumed across different device modalities~\cite{cross_device_lit_review}. Therefore, one of the core design rationale for \system was how the context-switching cost may be reduced and the relevant task context such as research threads may be surfaced as the end-user moves from one paper to another. On the one end, active reading interfaces such as \textit{LiquidText}~\cite{liquidtext} and \textit{texSketch}~\cite{texsketch} integrates an interactive canvas for note-taking and diagramming to an individual document to support in-depth reading. On the other end of the spectrum are systems for visualizing a collection of documents and their relevance between each other, as discussed in the related work section above. In contrast,} \threddy's PDF renderer seamlessly replaces the end-user's default PDF reader in their browser, such that they can read papers as they would normally without any additional constraint of having to start reading papers using a new system, \new{all the while collecting relevant papers and structuring them in a form that reflects the user's current mental model of the research space.}
\begin{figure*}[t]
    \centering
    \includegraphics[width=.95\textwidth]{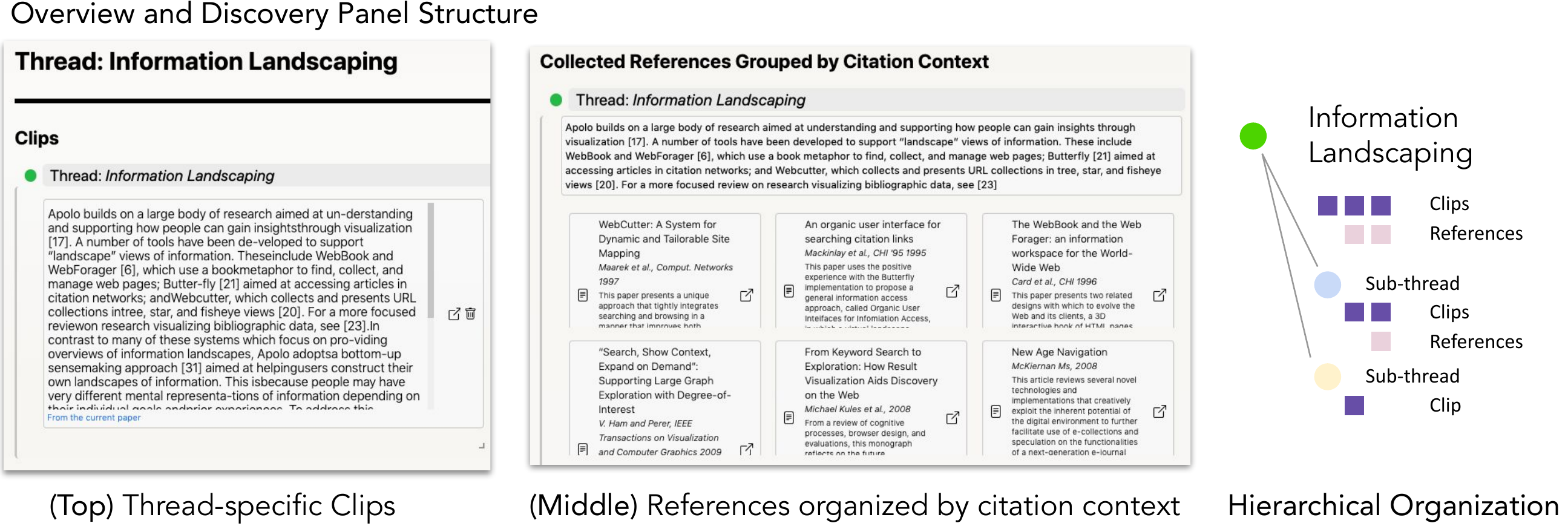}
    \vspace{-1em}
    \caption{The Overview and Discovery page consists of three components: At the top of the page is a section for the clips collected for the thread. In the middle of the page are references that belong to the thread grouped by their citation context, and at the bottom of the page (not shown here) is a  recommendation panel that contain relevant papers. The overview shows all of the entry thread and its sub-threads' content in a hierarchical manner (indented tree). Readers can choose which thread they want to look at in more details from the sidebar view.}
    \label{fig:overview}
\end{figure*}
\xhdr{Highlighting and Selection} \threddy's main PDF reader is divided into two areas (fig.~\ref{fig:main}): a PDF viewer and highlighter on the left \cirnum{E}, and a sidebar view holding the thread-related content on the right \cirnum{F}. The PDF viewer supports text \new{(using mouse drag)} and area highlights for images (drag-and-drop while pressing and holding the options/alt key), which trigger \system to extract references included in the highlighted context (\new{extraction is supported only for} text \new{highlights}). Readers can view the extracted citation context and references in the \cirnum{B} holding tank in the side bar, and deselect any reference they do not want to include or to fix any extraction error. Readers can add the citation context together with the references as a new thread, or simply add them as a clip or papers to an existing thread using the \cirnum{C} thread selector. The selector uses the citation context and computes the \textit{thread similarity} (Appendix A) to suggest which thread the highlighted and extracted content most likely belongs to. Threads and references are visually (e.g., \new{different threads} use colored dots with numbered counts of nested items on the left; papers use a `document' icon in place of the colored dots) and organizationally differentiated (e.g., papers show a distinct title - metadata - TL;DR \new{content} structure within each card UI). Readers can edit the context or the label in the thread by clicking on the text. Citation context clips are visible only in the Overview and Discovery by default, to prevent clutter.
\begin{figure*}[h!]
    \begin{center}
        \includegraphics[width=.75\textwidth]{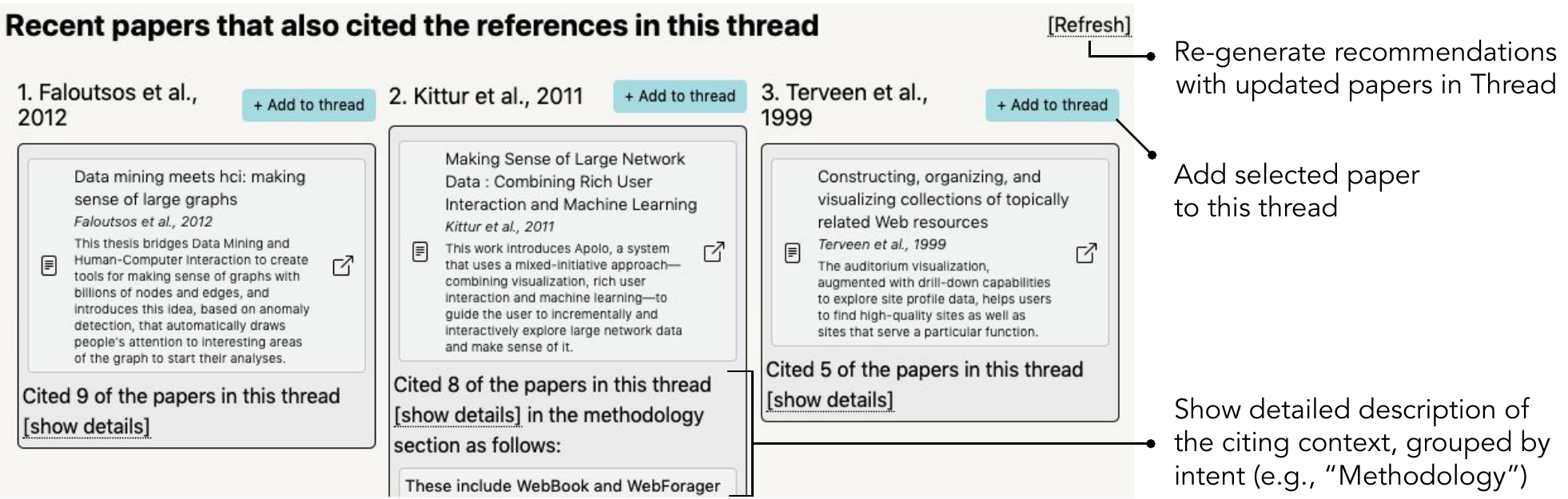}
        \vspace{-1em}
        \caption{The discovery view shows recommendations with high citation coverage, recency, and semantic similarity in a grid. Users can examine the details of each recommendation and decide to add the recommended paper to the current thread. Once the new paper is added, the user can click [Refresh] to generate new recommendations using the updated thread.}
        \label{fig:discovery}
    \end{center}
\end{figure*}
\xhdr{Organization} Readers can (re-)organize the threads by drag-and-dropping a thread or papers \textit{into} another thread as a nested thread or \textit{out of} a thread to start a new one. Threads that most recently received an additive change (e.g., a new paper was added to it) are moved to the top. The rationale behind this design choice was that readers may have more organizational needs for the threads that last received content, and/or they are the ones readers most recently attended to and thus are likely to be re-visited. At the top of the thread drawer \cirnum{D}, the \new{default} `Unorganized Papers' thread is created and each paper PDF opened in the viewer is \new{initially} added to it, such that readers can \new{re-visit or} organize it at a later time if they wish to (cf. `deferred actions' interaction design~\cite{hinckley2012informal}). The paper in the current reader is annotated with a `current paper' message at the bottom of the corresponding paper card \new{for awareness}.

\xhdr{Overview, Discovery and Persistence} Clicking on either the `View Details' or `Zoom' icon in each thread opens \new{the O}verview and \new{D}iscovery panel \new{for the selected thread} (fig.~\ref{fig:overview}). The clips that were minimized to prevent clutter in the sidebar view are now visible in full details, along with references grouped by the citation context they were collected from. The panel shows all of the nested threads and their content structured in a hierarchical manner, along with the content of the selected thread. \new{At the bottom of the Overview and Discovery panel are new paper recommendations generated by searching for those that have most cited the papers curated for the selected thread (fig.~\ref{fig:discovery}). Each recommended paper conveys relevance by showing the number of papers that it cited from the curated, along with the citation context and intent (e.g., in the `Methodology' section). These more recent papers help users discover newer development on the relevant research threads, akin to \textit{forward chaining} commonly used by scholars conducting literature reviews.}

\section{System Architecture}
\begin{figure*}[t]
    \includegraphics[width=.95\textwidth]{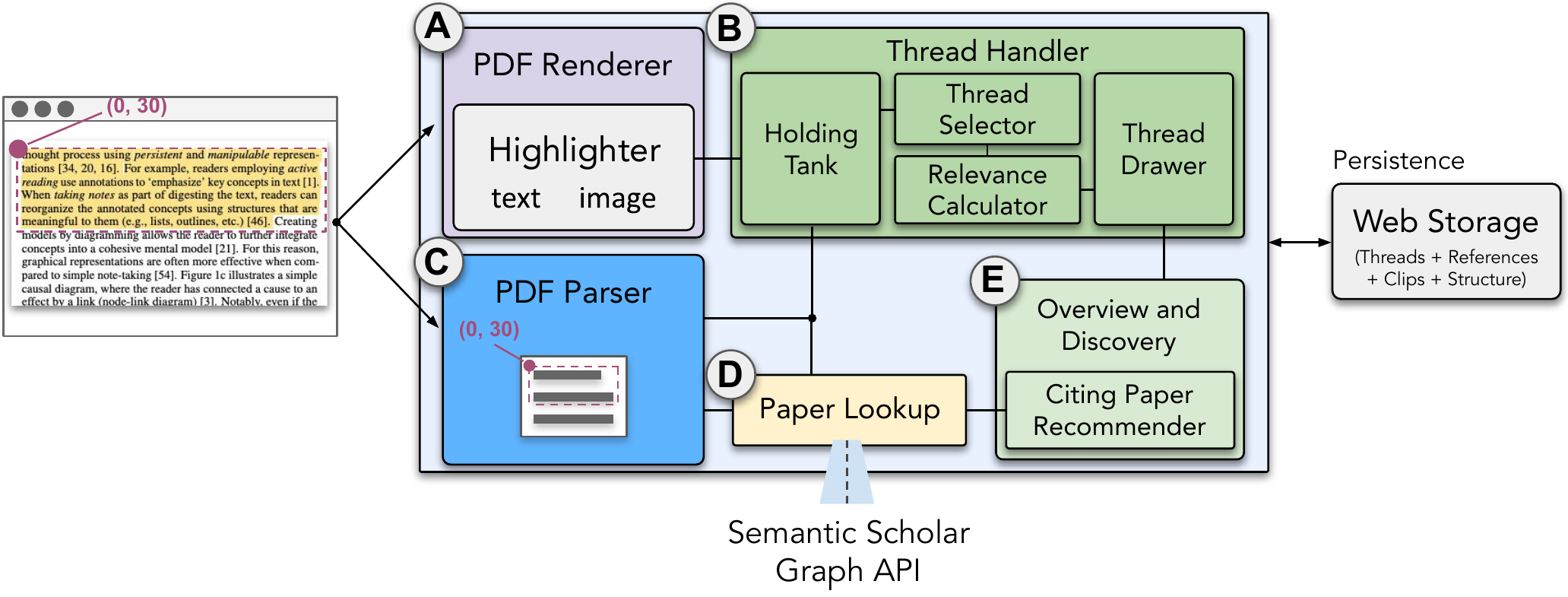}
    \vspace{-1em}
    \caption{\system system architecture: (A) PDF Renderer captures and handles user highlight events, (B) Thread Handler fetches the references included in the highlighted text and visualizes them in the holding tank. (C) The parser provides the PDF parse to the thread handler for finding corresponding references in the highlighted text and fetching them via the (D) Paper Lookup module. \new{It also supports (E) the citing paper recommender for user expanded threads.} The data is stored in a web storage and persisted as the reader moves to other papers.}
    \vspace{-1em}
    \label{fig:system_overview}
\end{figure*}

\new{The front-end of \system handles user interaction with PDFs, rendering of threads, and the Overview and Discovery panel and is implemented} as a Chrome browser extension. The back-end is implemented as a Flask server using GROBID~\cite{grobid} \new{for parsing the PDF content such as the title, section headers, each in-line citation notation to the corresponding reference entry, and body text sentences and their coordinates within the document from its top left-hand side corner}.

\subsection{Automatic Extraction and Linking References to User Highlights} 
\xhdr{Backend PDF parsing, linking, and mapping user highlight locations} When the user opens a new paper PDF on the browser, \system sends the file data to the backend GROBID server for parsing. \new{Our GROBID server uses a cascade of sequence labeling models including a fast linear chain CRF to parse a PDF (see~\cite{grobid} for more details)}. This server provides full text extraction and structuring of the received PDF, including the overall document segmentation (i.e., locating elements in pixel positions given the scale of the received PDF file) and structuring the text body into sentences, section titles, \new{in-line citation notation to corresponding references (e.g., Whether and which reference `[1]' in a sentence represents)}, figures, tables, etc. This process is run once when a new paper is opened in \system (previously processed PDFs are cached) and takes \new{up to} a few 10s of seconds to complete. Additionally, we include the pixel coordinates for each `sentence' parsed from GROBID, and run an additional sentence parsing using spaCy\footnote{\url{https://spacy.io/}} to \new{merge} sentence\new{s} that \new{may have been erroneously broken}. Using this parsed data, we search on the S2ORC~\cite{S2ORC} and Semantic Scholar APIs\footnote{\url{https://www.semanticscholar.org/product/api}} to link the corresponding paper with its metadata including the URL, publication year, TL;DR~\cite{scitldr}, SPECTER embedding~\cite{specter}. 

Using the parsed PDF, we \new{first} align the scale of the rendered PDF with the PDF used for parsing in GROBID. When the end-user highlights a portion of the text in PDF, the scale-adjusted coordinates of the highlight location is used to search overlapping sentence coordinates of the parsed PDF. We also collect the surrounding context (i.e., pre- and post-sentences of the overlapping sentence) for clipping. References included in this contextualized selection are searched in the parsed PDF. For image highlights, we simply take a screenshot of the underlying content and convert it into a data url for storage and display.

% \subsubsection{Data Persistence Across PDFs on the Web} \new{What is the best way to describe Schema here?}
% Chrome extension and Web user data storage

\subsection{User Highlights and Creation of Threads}

\xhdr{Front-end PDF viewer, Highlighter, and Sidebar} Using the parsed PDF data, we replace the native Chrome PDF viewer with our custom viewer based on an existing highlighter\footnote{\url{https://github.com/agentcooper/react-pdf-highlighter}} that provides convenient functionality for text and area selection which is a wrapper around the underlying rendering engine based on Mozilla's PDF.js\footnote{\url{https://github.com/mozilla/pdf.js}}. We feed it with our parsed PDF data and align the rendering scale and user mouse coordinates in accordance with the parsed PDF's coordinate system.

The sidebar consists of multiple components (fig.~\ref{fig:main}). The \textit{holding tank} view at the top visualizes the intermediate content based on user selection. This includes the user highlighted text and the references that are directly in or nearby the highlighted content or an image highlight from the PDF. The references are shown as a list of cards underneath the user highlighted content. Each reference card contains title and additional metadata about the paper, as well as the surface citation notation as shown in the PDF text for ease of mapping. The user may choose to select some but not all of the references automatically extracted to include (clicking on the trash icon discards the selected reference). The \textit{toolbar} underneath the holding tank is selectively activated based on user selection of the thread and the input data. The text input box (fig.~\ref{fig:main}, \cirnum{C}) allows users to either type in new text (will create a new thread if no matching thread exists), or select one of the existing threads to add the data to. Thread suggestions are generated using an algorithm that first compares user highlighted text and to each vertical chain of the threads to find the most related top-level thread. Next it ranks the most closely related sub-thread within the best chain that the new content may be added to (see Appendix A for details).

\xhdr{Thread interaction} Threads are presented using an interactive nested structure which users can drag-and-drop to nest a thread under another or move it out of the parent thread, delete an existing thread, modify the label of the thread, or collapse/expand all its content (fig.~\ref{fig:main}, \cirnum{D}). At the top of the thread view is a single-level, unremovable thread titled `Unorganized Papers' -- this thread is automatically generated and adds any paper PDF that is opened by the user under it, allowing the them to defer the action of organizing the papers. This thread is not subject to the user interactions described above other than moving its member references to a different thread (or vice versa). 

\noindent{\textbf{Clips and references}} included in each thread are visually differentiated. Clips are not shown to the users by default, but simplified as a simple counter message (e.g., `3 clips found. View details'), upon clicking which opens up an Overview and Discovery panel (Section.~\ref{subsection:overview}) to allow for the end-users to examine further details. In our user study, participants often clipped several textual and image content for each thread; showing all of them in the thread view will clutter it and make it hard to find the threads the user was building on. References, on the other hand, are directly shown as a list of separate cards underneath the thread that they belong to aid immediate access and further exploration. In addition to the content and metadata of the referenced paper, each paper card UI contains a URL icon, which automatically links the reference to its URL on Semantic Scholar. Clicking on the link directs end-users to the paper details page and reduces the amount of context switch they otherwise need in order to find the PDF.

\subsection{Overview and Discovery of Additional Papers Related to Threads}
\xhdr{Overview of threads} \label{subsection:overview} Once users have created multiple threads that may include sub-threads and relevant references, it can be challenging to review all of the collected and organized content, and to use all of its data to find additional papers in the related literature to further explore and grow the thread. To aid users with this overview and discovery experience, we designed a direct access to a separate panel (fig.~\ref{fig:overview}) which opens up when the user clicks either the `View details' text or the Zoom icon included in each thread card. The \new{panel is expanded to} the whole screen width when opened and shows the unrolled view of the selected thread and all of its sub-threads, along with the clips and references collected \new{at each depth}. We visualize this using an indented hierarchy to further differentiate threads at different levels (fig.~\ref{fig:overview}, right). The panel has three main sections: At the top, user collected clips are shown in a grid with accompanying annotations of the source they were clipped from. Next, a list of references is grouped by their citation context and presented. Finally relevant papers are recommended at the bottom of the page.

\xhdr{Discovering new papers} \label{subsection:discovery} Using the data collected and organized by each thread to find more relevant papers to further grow the thread could be a challenging task and may significantly interrupt with the end-user's flow of reading. In order to close this discovery loop, we automatically recommend new relevant papers \new{when the Overview and Discovery panel is opened}. End-users can \new{then} select any of the returned recommendations to add to the thread by clicking on the `Add to thread' button (fig.~\ref{fig:discovery}, middle). This adds the paper to the thread in the sidebar as well and keeps it in sync. End-users may click `Refresh' (fig.~\ref{fig:discovery}, top right) to re-generate the recommendations based on the updated list of references in the thread, which then invokes the recommender engine.

We use the Semantic Scholar API\footnote{\url{https://www.semanticscholar.org/product/api}} to fetch necessary paper details for retrieval. 
%While the paper citations information is available via the API in batches (up to size 1,000), the detailed information including paper embeddings (\textsc{Specter}~\cite{specter}) and \textsc{tl;drs}~\cite{scitldr} are searchable on an individual paper basis at the point of submission of this paper. 
Our recommendations use \textit{citation coverage} as its primary source of relevance signal. The rationale behind this decision is that each \new{selected} thread contains user\new{-curated} relevant papers and the higher the number of thread references cited by a new paper, the higher the chance that it may be relevant to the thread. While we limit our \new{search boundary} up to 1,000 direct citations for each thread reference, future work may explore relevance via longer citation chains. For each of the citing paper, we \new{simply count how many of the unique} thread references were cited by the new paper. In our pilot tests, we found this to be a good proxy for relevance to the thread's content and return to this in our discussion. We sample a much smaller number (50) of top-ranked results from the top-ranked high-coverage citing papers and sort them by their publication recency. If the two citing papers have the same publication year, we further differentiate them by their semantic similarity computed as the cosine similarity between the centroid vector of the set of thread references \new{included in the thread} vs the new citing paper using SPECTER embeddings.

\section{Evaluation}
In evaluation our goal was to study how effectively \system supports scholars reading research papers to review the relevant literature in a new domain. To this end, we designed a short literature review task with the goal of producing an outline structure either for themselves in the future or someone else to build upon. We employed a within-subject study comparing \system to the commonly used \textsc{Google Docs} editor baseline \new{(without any extensions for searching research papers installed)} with research topics that scholars were personally interested in conducting a literature review of. Alternative choices for the baseline comparison may include combining a qualitative coding software such as NVivo with a reference management tool Mendeley\footnote{\url{https://tinyurl.com/yckre5md}}, or the recently announced Zotero 6\footnote{\url{https://www.zotero.org/blog/zotero-6/}} \new{browser plug-in} which allows users to annotate PDF documents opened in it and create notes \new{that can be exported and imported into the desktop Zotero application}. There are pros and cons of choosing each of these alternatives as a baseline for our comparison. However, in our study we decided to use \textsc{Google Docs} because: a) every participant currently uses it or has used it before to conduct literature review; b) it was directly accessible to everyone; and c) it was sufficiently versatile to support creation of research threads, clipping, and adding references to the thread. We return to the choice of the baseline condition in Discussion.

\xhdr{Participants and process} We recruited 9 participants (1 female) for the study. We employed a within-subjects study design, and counterbalanced the order of presentation using 4 Latin Square blocks and randomized rows. Due to the uneven number of recruited participants, the \textsc{Google Docs}-first presentation order was assigned one more time than the \system-first order. The mean age of participants was 29.3 (SD: 4.67) and all actively conducted research at the time of the study (1 Master's student, 2, Post-docs, 6 PhD students). \new{Participants' fields of studies included: HCI (5), NLP (2), Material Sciences (2).} Participants followed the following process in the study\new{, which took place remotely using Google Meets}: introduction and consent, installation of Threddy, two training tasks followed by the main tasks in an individualized order, and surveys. \new{Participants were asked to share their screen during the study.} We ended the study with a debrief interview with participants in which the interviewer asked follow-up questions on his observations. The study lasted around 1 hour 20 minutes and participants were compensated at a \$30 USD per hour rate.

\xhdr{Training tasks} We used the following paper~\cite{paragon} and using one of its subsections in the Related Work (4 paragraphs) \new{as the seed for practicing creation of} an outline. The experimenter described the concepts used in the main task \new{including}: `Threads', `Clips', and `References'; Threads are short descriptions of topics or concepts in the related domain and can form a hierarchy with other threads; Clips are supporting pieces of information related to a thread, which can range from a phrase to a paragraph-length text or images directly taken from the paper; References are papers relevant to the thread. Participants were shown a simple example outline and instructed that the outline needed a sufficient amount of details for comprehension and the \new{accessibility of the source}. In addition, participants were instructed to read at least one more paper that is relevant and create an outline that can incorporate multiple references in it, starting from the seed paper. Participants were shown a quick tour and core functionality demonstration (5 minutes) followed by a task to recreate the outline they created in a Google Doc using the same text in Threddy.

\xhdr{Timed main tasks} The main tasks used the two topically diverse papers that participants submitted as personally motivating sources for their own literature reviews as part of the sign up process. We randomly assigned each paper to a condition and instructed the participants to start from it as a seed for the task. The tasks were performed for 20 minutes each. 

\xhdr{Surveys and interview}
For demand (including physical and cognitive) and overall performance we adopt the validated 6-item NASA-TLX scale~\cite{nasa_tlx}. For technological compatibility with participants' existing literature review workflows and the easiness of learning we adapted the Technology Acceptance Model survey from~\cite{tam_survey} (5 items). For measuring the flow aspect~\cite{csikszentmihalyi1990flow} of participants' interaction with the system, we adopt Webster et al.'s research~\cite{webster1993dimensionality} uncovering multiple interrelated dimensions of flow in human-computer interaction and the corresponding questionnaire (11 items). Finally, we included 8 additional questions asking participants about extraction of clips and references, as well as their organization into a thread structure \new{(See Appendix B for details of the questionnaire)}.

\xhdr{Coding} For the baseline, two of the authors coded the first participant's outline together to count the number of threads created, as well as the number of clips and references collected. Then the coders independently coded the rest of the data. The ordinal Krippendorff's alphas were significant for all categories: 0.842, 0.962, 0.822 for threads, clips, and references, respectively. The main sources of disagreement included: whether to count unfinished notes and incomplete text as clips, repeated or slightly modified paper title text as threads, and so on. The final sets of counts for the baseline condition were therefore produced by resolving any disagreements by taking the average between the two coders.

\section{Findings}
\begin{figure*}[t]
    \begin{subfigure}[t]{.25\linewidth}
        \centering
        \includegraphics[height=3.75cm]{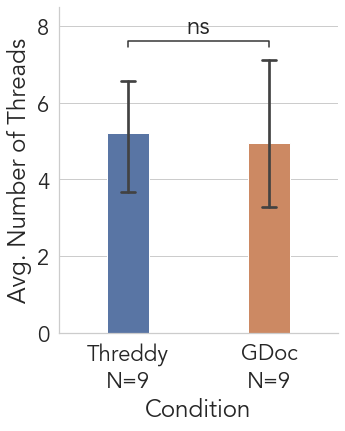}
        \caption{The number of threads participants created in \system ($\mu = 5.2, \text{SD} = 2.28$) and \baseline ($\mu = 4.9, \text{SD} = 3.03$) did not differ significantly (paired t-test, $t(14.87) = -0.22, p=0.8$).}
        \label{fig:threads}
    \end{subfigure}
    \quad
    \centering
    \begin{subfigure}[t]{.33\linewidth}
        \centering
        \includegraphics[height=3.75cm]{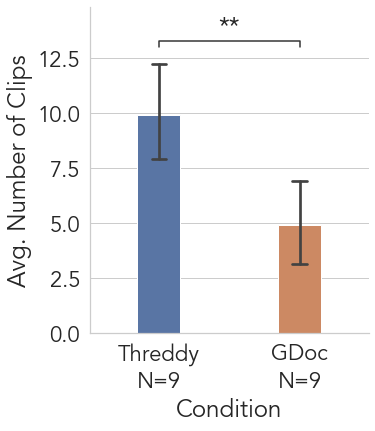}
        \caption{However, participants added significantly more clips to threads ($\mu = 9.9, \text{SD} = 3.48$) in \system vs. \baseline ($\mu = 4.9, \text{SD} = 3.17$) (paired t-test, $t(15.86)=-3.19, p = 0.006$) }
        \label{fig:clips}
    \end{subfigure}
    \quad 
    \begin{subfigure}[t]{.33\linewidth}
        \centering
        \includegraphics[height=3.75cm]{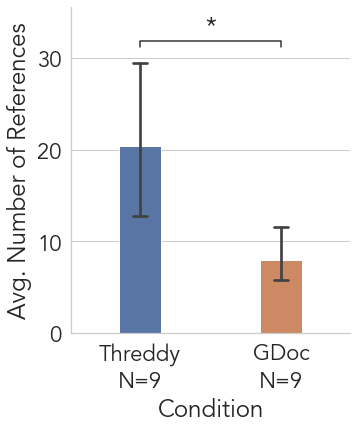}
        \caption{Participants also collected and placed a significantly higher number of references into threads ($\mu = 20.4, \text{SD} = 13.84$) in \system vs. \baseline ($\mu = 7.9, \text{SD} = 4.91$) (paired t-test, $t(9.98) = -2.55, p = 0.03$).}
        \label{fig:references}
    \end{subfigure}
    \vspace{-1em}
    \caption{Evaluation Results: (a) The number of threads created by participants during the task did not differ significantly between \system and \bs. However, participants collected and organized a significantly higher number of (b) clips and (c) references into the threads on \threddy.}
\end{figure*}

\subsection{Collection and Organization into Threads}
\subsubsection{Quantitative Results} Participants in both conditions created a similar number of threads ($\mu=5.2, \text{SD}=2.28$ in \system vs. $\mu = 4.9, \text{SD} = 3.03$ in \bs, paired t-test $p=0.8$). However, the number of clips collected for threads was twice as high in the \system condition ($\mu = 9.9, \text{SD} = 3.48$) than in the \baseline condition ($\mu = 4.9, \text{SD} = 3.17$, $t(15.86)=-3.19, p = 0.006$), demonstrating the utility of \system for supporting collection of clips for individual threads. In addition, the number of references collected and organized by the relevant threads was 2.6$\times$ higher ($\mu = 20.4, \text{SD} = 13.84$) in the \system condition than in the \baseline condition ($\mu = 7.9, \text{SD} = 4.91, p = 0.03$), further demonstrating its support for efficient collection and organization of references by their relevant threads.

\subsubsection{Qualitative Results} 
\hfill\\
\noindent\textbf{Saving time.} All of the participants mentioned that automatic extraction of references from the highlighted citation context and linking them with metadata saved time. P1 mentioned that it ``\textit{saves a lot of time because I don't have to cross check between the context and the references section}'' and similarly P3 said ``\textit{Collecting references is such a pain, such a context break... and when I go to the references section and finally connect the number to the actual paper, and it turns out that the paper itself doesn't even sound interesting, I need to go back to where I was, with the damage already being done in terms of breaking my reading flow.}'' Compared to their experience with \threddy, participants described their typical workflow of conducting literature review as involving lots of ``\textit{scrolling back and forth}'' and ``\textit{pointing and clicking}'' (7/9 participants), and having to switch between different applications such as search engines, PDF viewers, and note-taking applications (3/9). Automatic linking to metadata such as the link to the paper details page on Semantic Scholar allowed participants to do without ``\textit{having to spend a lot of time to track down PDFs}'' (P5, P7). Automatically binding references to their citation context ``\textit{reduced a few clicks that would have otherwise been necessary to organize and keep track of that way}'' (P9). Uniformly formatting references in the card UI removed the subtask for ``\textit{formatting the references, for example it's always a pain to format the text I copied from a PDF... the text is either too large, colored differently, or have weird line breaks, and often times the URLs are way too long}'' (P1) and made it easy as to not having to ``\textit{worry about all different (surface) citation forms}'' (P5). This saved time was thought to be used for reading additional papers or switching to a different thread to explore more (P2).

\xhdr{Context awareness and flow} Participants also described the persistence of threads and pinning most recently worked on threads to the top as effective awareness mechanisms for continuing their thinking along those threads. P9 said: ``\textit{Persisting data is helfpul, I don't have to go back to the previous paper to be reminded of what I was thinking of.}'' Context clipping was also considered helpful (3/9) and even better with its persistence across papers (P1). P9 thought being able to see the threads while reading any paper was ``\textit{a good forcing function to encourage myself to structure and organize as I go.}'' Finally, P5 described the benefit of this awareness as follows:
\begin{quote}
``Perpetualness of the information is nice because otherwise I have to do this kind of tasks in discrete chunks, just because I'll lose the context a lot in the process because I cannot see where things came from. Or just with a long list of citations and other things... I cannot remember exactly why something that I opened up later is relevant anymore. In comparison, here in the interface I have the continuous stream of cognition, like `Past Me thought this was relevant to [production of materials] (thread created by P5).''
\end{quote}
These results were also corroborated by participants' responses to survey questions. On a 7-point Likert scale (1: Strongly disagree, 7: Strongly agree), participants' agreement was significantly higher with the statement ``\textit{It was easy to collect relevant clips using the system.}'' in the \system condition ($\mu = 6.2, \text{SD} = 0.67$) than in the \baseline condition ($\mu = 5.2, \text{SD} = 1.09$, $t(13.23) = -3.46, p=0.01$). Furthermore, participants felt like it was easier ``\textit{to keep track of relevant references}'' using \system ($\mu=5.9, \text{SD}=1.48$) than \baseline ($\mu = 3.8, \text{SD} = 1.27, t(15.63) = -2.80, p= 0.02$), and also easier ``\textit{to organize references into relevant threads}'' using \system ($\mu = 5.6, \text{SD} = 1.13$) than \baseline ($\mu = 3.4, \text{SD} = 1.33, t(15.58)=-3.59, p=0.007$). Finally, participants also felt that they were in a heightened flow state while using \system based on the results from our survey. Participants' average composite responses to the flow questionnaire items was significantly higher in \system ($\mu = 51.0, \text{SD} = 6.5$) than \baseline ($\mu = 42.9, \text{SD}=6.77, t(15.97)=-2.94, p=0.02$). \new{Additionally, the overall demand required for accomplishing the tasks, measured as the sum of the scores to five NASA-TLX questionnaire items (excluding performance), showed no significant difference between the \baseline ($\mu=62.6$, SD=25.90) and \system conditions ($\mu=55.3$, SD=20.34) ($t(15.15)=0.76, p=.52$, Two-sided paired samples t-test; Table~\ref{table:full_survey}).}

\subsection{Discovering Papers Relevant to Threads}
Participants mentioned that ``\textit{it was nice to see recommendations directly relevant to each thread}'' (3/9) and directly accessible in the interface (8/9). Many recommendations ``\textit{seemed relevant}'' (7/9) and especially in the first few rows of the recommendation section. Given their relevance, P6 felt that ``\textit{citation coverage as proxy for relevance seems to work well.}'' Participants also agreed with the statement ``\textit{It was easy to find additional papers relevant to each thread using the system}'' significantly more in the \system condition ($\mu = 6.1, \text{SD} = 0.78$) than in the \baseline condition ($\mu = 4.2, \text{SD} = 2.28$, $t(9.86) = -2.35, p=0.04$). They also felt featuring recommendations in terms of more recent papers that cited the references included in threads was helpful for ``\textit{getting a sense of how the field is progressing}'' (P3, P5). Interestingly, one participant felt that she experienced ``\textit{no fear of missing out (FOMO)}'' (P1) given the amount of recommendations available and their overall relevance, but another participant commented that ``\textit{I feel a little bit of FOMO because I built a thread which I thought was complete in a sense, but then seeing all these interesting articles made me think, what other things have I missed or simply not shown to me because I decided to organize this thread as such}'' (P5).

\subsection{Extracting Pre-digested Threads, Integrating Papers into Assembled Threads}
One challenge with extracting pre-digested threads from other papers was \textit{merged context}, which happened when the author of the paper combined multiple kinds and levels of research contribution into a single patch of text. P1 mentioned that ``\textit{(fuzzy) selection is nice, but it sometimes leads to too much... I have to still go through the list of extracted results and `check off' something that I don't want. For example, these papers are general and maybe related in the loose sense at best... these on the other hand are more specific.}'' Similarly P5 commented: ``\textit{My main frustration is that people put so many references into a single sentence, and they are not the same. Some of them are more specific and some of them are more general.}'' Participants mentioned a need for specifying when they want an exact or pin-pointing selection mechanism that complements the current fuzzy extraction from the highlighted citation context (``\textit{Sometimes I want to point at exactly one cite I want to add from the context\new{.}}'' -- P1), with additional support when a number of references were included in the citation context (``\textit{And for other papers, (citation) styles like [12 -- 15] are not great because they (the references) are in a batch, so how am I supposed to know which ones go to which and which one's interesting?\new{''}}). While automatically linking the metadata of the extracted references including the title and TL;DRs was helpful to the participants, sometimes they were unavailable due to missing data or provide insufficient context for comprehension specifically related to the citation context of interest (P1, P2, P5).

\section{Discussion}
\subsection{\new{Impact on the Workflow}}
\xhdr{\new{Thread-first vs. paper-first exploration}} Participants used \system to navigate through authors' pre-digested threads, collect ones that were interesting to them, and assembling their own threads using the collected threads. Compared to how they conducted literature review \system seemed to unlock a new capability for them to synthesize along the personally interesting threads, across multiple papers, without losing context in the process. P1 contrasted her experience on \system with how she currently conducts literature review on \baseline and described that the former inverses the role between threads and papers in a sense:
\begin{quote}
    ``On \bs, I can only focus on one paper at a time. When I find an interesting new paper, I'll skim it and write a short description of why it's interesting. If I find it's worth a more detailed read, I'll try to read it in full, taking notes... like creating my version of an annotated bibliography. In \system I can see my threads when I read a new paper, so I almost focus on those (threads) as opposed to individual papers.'' -- P1
\end{quote}
This phenomenon of \textit{thread-first} exploration (in contrast to the \textit{paper-first} exploration commonly used by our participants) was also observed in how our participants engaged with actions such as adding new papers to threads, moving references and sub-threads out of their parent thread once enough references with sufficiently different context were identified, and re-labeling and re-framing threads with different text to capture their evolving understanding of the research field. \new{It seemed that \system did not create additional workload for users by encouraging them to take the thread-first perspective either; there was no significant difference in terms of the number of threads participants created within the duration of the timed study in each condition nor in terms of their perception of the demand required of them. It is possible, however, that the core of the literature review task that requires complex processing of information (e.g., identifying interesting research threads, summarizing and labeling threads, organizing them in a useful structure, updating the threads) is not helped directly by \threddy. Indeed, participants perceived the easiness of creating threads and subthreads to be roughly equal in both \baseline and \system conditions (see Table~\ref{table:full_survey}). Instead, \system may help scholars with the task by freeing up their capacity and attention span that would otherwise be tied up by significant auxiliary tasks such as collecting supporting pieces of evidence or relevant context as clips, organizing them into threads, and growing the threads with additional references while keeping track of them. Indeed our participants responded that \system provided significant support for such tasks (Table~\ref{table:full_survey}).}

\xhdr{\new{On-the-go foraging and structuring}} \new{At a high level, \system users could continuously collect information while following the relevant threads of research. For example, improved context awareness and persistent threads across papers led participants to move between them and ``\textit{structure information on the go}'' (P9) as opposed to reviewing papers individually and ``in discrete chunks due to the frequently lost context'' (P5). Moreover, in-thread recommendations integrated the stages for searching and reading, further reducing the context-switching cost. These reduced switching costs and support for externalizing working memory position \system to be especially useful in supporting the early sensemaking process of literature review as researchers forage for information, helping them create what Pirolli and Card term a ``shoebox'' of relevant information~\cite{pirolli2005sensemaking} easily, in a more organized way, and with affordances for helping them pull in even more relevant information. There are many other aspects of the process where other tools would remain useful. For example, active diagramming and concept mapping may help users externalize representations focused on relations of the concepts involved (e.g., understanding how the different components of a weather system fit together from a paper, rather than the threads of research on weather systems). Synthesis and summarization tools may also help further along the process; here it is possible that a thread-based approach could scaffold the creation of synthesized mental models by enabling users to work with pre-grouped sets of papers.}

\xhdr{\new{Checking assumptions via thread-centric recommendations}} Participants generally appreciated the recommendations specific to each thread. In addition to the primary benefits of: 1) seeing what is out there, 2) getting a sense of how people are building off of the work curated for each thread, and 3) what may be relevant papers for further exploration along the specific threads, recommendations had secondary, unanticipated benefits as a ``check for whether my thread makes sense (by looking at the returned recommendations)'' (P3) and as a way to ``think about how I might re-define or sub-divide my threads'' (P7). Other participants felt additional mechanisms for specifying which context is personally more important for the recommendations, because they felt like ``the list of recommendations is quite a spoonful (of papers), some of them are relevant but only at a high level, for example I don't want to see [this paper] just because most of the references in my thread have cited it in the background'' (P1). A fruitful avenue for future work therefore may lie in designing alternative mechanisms for finding relevant papers to recommend for each thread that go beyond the simple citation coverage metric explored in this work.

\subsection{\new{Scaling over a Long Period of Time}} \new{An open question with Threddy is how the system would scale over time. With continued use, the number of threads and papers in them would grow significantly. Furthermore, a user's organization would require refactoring as they become more expert in an area; research areas grow and split; or their interests and mental models change. The design of Threddy was directly motivated by issues with scale faced by researchers using other collection tools such as Zotero or Mendeley, with the introduction of hierarchical threads developed in-situ aimed at providing a flexible and scalable way to keep track of diverse topics and subtopics during literature review. Though limited by the duration, our user study uncovered preliminary results speaking to the challenges of scale and time.}

\new{First, we found that users desired to focus primarily on a relatively small number of active threads that were most relevant to the context of their target paper. During the post-study interview participants noted that they are often limited in time by deadlines and focused on compiling the most relevant literature for their papers or grants, for example writing a related work section that might include 2-4 topic threads. As participants grew these threads, they noted that their mental models changed with new information, and could use Threddy to refactor the threads appropriately. This included pulling out a particularly dense topic into new subthreads, renaming threads as they learned more about what actually went in them, and using the hierarchy to nest new threads into existing ones and move between them.}

\new{However, participants also noted challenges such as not being able to see an overview of all their threads and easily reorganize them within a dedicated workspace. These suggest clear areas for future work including bringing in proven triage and workflow approaches such as tracking which papers have been read or are in different states of processing beyond the simple recency based mechanism introduced here. While these techniques  were not core to testing the thread-based idea but they will become essential to a real-world system involving many threads, clips, and references. Another area includes support for working with threads over time, including more intelligent ways to split and merge threads or reorganizing them in the thread hierarchy, which would likely become more important as a user's library grows over time. At an even higher level, there are interesting questions around whether \threddy's hierarchical structure might be improved on by more flexible graph structures, and how such representations could be collaboratively aggregated and built on by others.}

\vspace{-2em}
\new{\subsection{Beyond Citation Chaining} Chaining the references in forward and backward directions in time is a common practice used by scientists searching for high relevance papers in the literature and making sense of how the field has progressed over time~\cite{watkinson2016changes}. However, one potential limitation of citation chaining-based approaches is that it may limit the discoverability of work outside frequently co-cited bodies of literature, and may lead to filter bubbles~\cite{filter_bubbles_www_recommender}. Certain domains of knowledge are less likely to interact with each other~\cite{swanson_undiscovered,chu2021slowed} despite their potential for catalyzing significant scientific innovations~\cite{rzhetsky2015choosing}.  Here, we believe that augmentative tools that help end-users discover articles directly in the context of their flow of reading have potential for helping scholars become more open to the literature that may exist outside the domains they are familiar with but are nonetheless relevant. Additionally, the reduced context switching with the aid of the tool may also help scholars more deeply engage with more distant articles.}

\new{In this vein, recent work on discovering analogical scientific literature~\cite{kang_augmenting_tochi} demonstrated an early evidence of the feasibility of computationally sourcing analogical papers that, although may be missed by conventional search engines, would inspire scientists to come up with novel conceptualization of their research problems (see also~\cite{hope_scaling} for how similar mechanisms of sourcing computational analogies may spark inspirational ideas in a different task context). Complementary approaches may leverage the knowledge domains of papers that the scholar has recently read to automatically increase the frequency of cross-domain retrieval in subsequent recommendations (cf.~\cite{naacl2022_kang_augmenting}) or to design a user control for interactively tuning the retrieval domain diversity. At an even higher level, a broader design space for opportunities exists, for example how the system might source recommendations by taking into account the inferred `social' relevance to frequently read or cited authors, or use such relevance for user engagement and prioritization of recommendations~\cite{kang_from_who}.}

\section{Conclusion}
In this paper we developed \threddy, a system that supports users with collecting patches of text that contains pre-digested synthesis by other authors (i.e., useful citation context along with automatically extracted associated references), and helps them assemble threads they are personally interested in using clippings of other authors' pre-digested threads included in the introduction or related sections of papers. In contrast with prior work that sought to create separate information environments for similar objectives, \system seamlessly integrates into the user's in-situ context of reading, and aims at reducing the cost of context switching while harvesting, assembling, and synthesizing research threads. Further research is required to uncover additional design implications for in-situ reading support for collecting other's synthesis work and assembling them into their own threads.

%%
%% The acknowledgments section is defined using the "acks" environment
%% (and NOT an unnumbered section). This ensures the proper
%% identification of the section in the article metadata, and the
%% consistent spelling of the heading.

\begin{acks}
This work was supported by the Carnegie Mellon Center for Knowledge Acceleration, National Science Foundation (FW-HTF-RL, grant no. 1928631; IIS, grant no. 1816242; SHF, grant no. 1814826), The Allen Institute for Artificial Intelligence (Semantic Scholar), Google, and the Office of Naval Research. We also thank the anonymous reviewers for their constructive feedback. We also extend a heartfelt thanks to our study participants without whom this work would not have been possible.
\end{acks}

%%
%% The next two lines define the bibliography style to be used, and
%% the bibliography file.
\bibliographystyle{ACM-Reference-Format}
\bibliography{main}

\newpage
\subsection*{Appendix A. Algorithm for Ranking Relevant \new{Existing} Threads for \new{Adding New Threads}}
\vspace{-1.5em}
\label{appendix:algo_thread_relevance}
\begin{figure}[h!]
    \includegraphics[width=.45\textwidth]{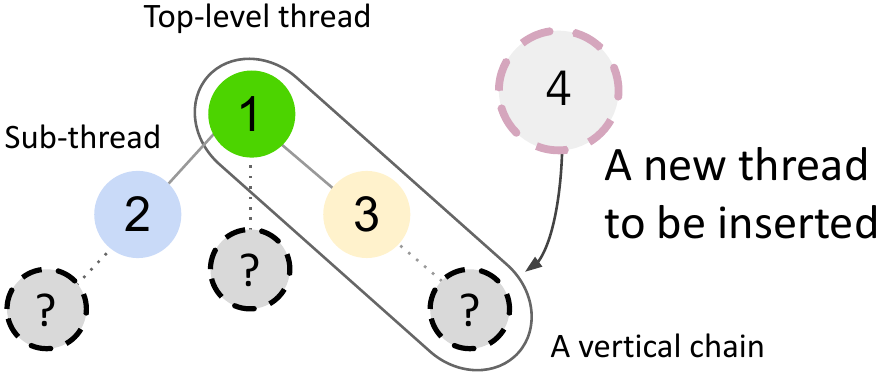}
    \vspace{-1em}
e    \caption{Closely related threads given a new target thread to be inserted are \new{ranked in two stages.} In stage one, the vertical chains of threads are \new{grouped together}, and a measure \new{of fit that balances} the group similarity (i.e., similarity to the centroid of the group) and the maximal member similarity is computed. \new{In stage two, each of the member threads of a vertical chain is compared against to the target thread to further rank them based on similarity which helps the end-user quickly see which thread may be the most relevant for association. When no thread is matching the new thread's content the user can insert it as a new top-level thread.}}
    \vspace{-1.5em}
    \label{fig:thread_rec}
\end{figure}

\new{With the continued use, scholars would likely accumulate a number of threads in the drawer of the interface, leading to a  scanning cost that may increase linearly with it at the minimum. In the hope of scaling the usage of \threddy, an algorithm was developed to automatically sort} closely related threads that the new thread most likely belongs to \new{in a descending order of their relevance.} \new{The first step of this algorithm is finding} closely related \new{vertical} `chains' of threads to the to-be-added thread. The intuition here is that end-users most likely nested threads for their semantic relatedness, preserving of which may provide the system a valuable source of signal for discerning the similarity between the new thread and \new{existing chains of} threads. Therefore, we \new{first} group the members of each chained threads \new{(traversed via the depth-first manner)}, and compute the similarity between the new target thread and the centroid of the chain. \new{For computing the similarity and the chain centroid,} the target thread and each member of the group are embedded into high-dimensional vectors that preserve their multifaceted semantic relatedness. We use the Microsoft’s MiniLM model~\cite{wang2020minilm} fine-tuned by HuggingFace\footnote{\url{huggingface.co}} with 1B+ training pairs, including 116M citation pairs from S2ORC. In our pilot test, this model provided a good trade-off between efficiency and performance for use in our real-time application setting. \new{The chain centroid is computed using a simple average of the member thread embeddings.} 

\new{However,} optimizing only for the similarity to the group centroid runs the risk of finding a \new{chain} that although \new{the} members' centroid is close to the target thread's embedding, all of its members \new{may be scattered} far from one another (\new{i.e.,} high dispersion, low cohesion). Therefore, we further measure the similarity between the target thread and the closest member thread in the chain and use it \new{to deprioritize matching on such cases:}
\begin{align*}
    \text{Group Similarity}_{(\text{grp}, \text{T})} &:= \text{sim}\left(\sum_{n \in \text{grp}}^{N}(\overrightarrow{\text{emb(n)}}) / N, \overrightarrow{\text{emb(T)}} \right) \\
    \text{Cohesion}_{(\text{grp}, \text{T})} &:= \text{max}_{n \in \text{grp}}\text{sim}\left( \overrightarrow{\text{emb(n)}}, \overrightarrow{\text{emb(T)}}\right)
\end{align*}
For a given target thread $T$, our final rank objective is multiplicative:
\begin{align*}
    \text{argmax}_{\text{grp}}\left(\text{Group Similarity}_{\text{grp}, T} \times \text{Cohesion}_{\text{grp}, T}\right)
\end{align*}
to prioritize groups with coherent rather than lopsided similarity (e.g., a high score on only one of Group Similarity or Cohesion but low score on the other may result in an overall irrelevant thread to the user \new{due to the potential situations as described above.}). Once we have identified the best chained thread to insert the target thread into, we further rank its member threads in the order of its similarity to the target thread embedding. \new{The resulting ranks of the threads are then presented to the user who may insert the target thread at a particular position of the chain.}

\subsection*{\new{Appendix B. Additional Survey Results}} \label{appendix:survey_results}
\new{Descriptions of additional questionnaire items and participants' responses grouped by condition are presented in Table~\ref{table:full_survey}. Two-sided paired samples t-tests were performed to compute the $p$-values between conditions. See Section 7.1 for a discussion of the results.}
\begin{table*}[t!]
    \centering
    \begin{tabular}{p{2.75cm} p{7.4cm} p{2.25cm} p{2.25cm} p{1.35cm}}
    \toprule
    & \textbf{Description} & \bs & \system & \textit{p}-val. \\
    \midrule
    \multirow{2}{*}{{Overall Workload}} & Sum of the participants' responses to the five NASA-TLX's~\cite{nasa_tlx} 21-point scale questionnaire items below. & \multirow{2}{*}{62.6 (SD=25.90)} & \multirow{2}{*}{55.3 (SD=20.34)} & \multirow{2}{*}{$p=.47$} \\
    \midrule
    Mental Demand & ``How mentally demanding was the task?'' & 10.4 (SD=6.91) & 11.6 (SD=5.59) & $p=.60$ \\
    \addlinespace[.1cm]
    Physical Demand & ``How physically demanding was the task?'' & 12.7 (SD=7.16) & 9.3 (SD=4.18) & $p=.14$ \\
    \addlinespace[.1cm]
    Temporal Demand & ``How hurried or rushed was the pace of the task?'' & 13.1 (SD=6.15) & 12.3 (SD=4.47) & $p=.64$ \\
    \addlinespace[.1cm]
    \multirow{2}{*}{Effort} & ``How hard did you have to work to accomplish your level of performance?'' & \multirow{2}{*}{16.2 (SD=3.90)} & \multirow{2}{*}{11.7 (SD=5.39)} & \multirow{2}{*}{$p=.15$} \\
    \addlinespace[.1cm]
    \multirow{2}{*}{Frustration} & ``How insecure, discouraged, irritated, stressed, and annoyed were you?'' & \multirow{2}{*}{10.1 (SD=6.72)} & \multirow{2}{*}{10.4 (SD=5.94)} & \multirow{2}{*}{$p=.92$} \\
    \midrule
    \multirow{3}{*}{{Flow}} & Sum of the participants' responses to the 11 questionnaire items adopted from Webster et al.~\cite{webster1993dimensionality} measuring the flow aspect of participants' interaction with the system. & \multirow{3}{*}{42.9 (SD=6.77)} & \multirow{3}{*}{51.0 (SD=6.50)} & \multirow{3}{*}{$p=.02^{*}$}\\
    \midrule
    \multirow{4}{*}{{TAM}} & Sum of the participants' responses to the 5 questionnaire items adopted from~\cite{tam_survey} measuring the technological compatibility with participants' existing literature review workflows and the easiness of learning. & \multirow{4}{*}{28.1 (SD=5.93)} & \multirow{4}{*}{24.7 (SD=5.12)} & \multirow{4}{*}{$p=.18$} \\
    \midrule
    \multirow{3}{*}{Confidence} & ``Using the system increased my confidence in conducting literature review. (The response Likert scales for this question and below are 1: \textit{Strongly disagree}, 7: \textit{Strongly agree})'' & \multirow{3}{*}{4.7 (SD=1.87)} & \multirow{3}{*}{4.8 (SD=1.56)} & \multirow{3}{*}{$p=.88$} \\
    \addlinespace[.1cm]
    \multirow{2}{*}{Creating Threads} & ``It was easy to create different threads in the related literature using the system.'' & \multirow{2}{*}{5.6 (SD=1.24)} & \multirow{2}{*}{5.9 (SD=0.93)} & \multirow{2}{*}{$p=.54$} \\
    \addlinespace[.1cm]
    Creating Sub-threads & ``It was easy to add sub-threads using the system.'' & 5.4 (SD=1.51) & 5.3 (SD=1.73) & $p=.83$ \\
    \addlinespace[.1cm]
    Collecting Clips & ``It was easy to collect relevant clips using the system.'' & 5.2 (SD=1.09) & 6.2 (SD=0.67) & $p=.001^{**}$ \\
    \addlinespace[.1cm]
    \multirow{2}{*}{Organizing Clips} & ``It was easy to organize clips into relevant threads using the system.'' & \multirow{2}{*}{5.8 (SD=0.97)} & \multirow{2}{*}{5.3 (SD=1.50)} & \multirow{2}{*}{$p=.27$} \\
    \addlinespace[.1cm]
    Keeping Track of References & ``It was easy to keep track of relevant references using the system.'' & \multirow{2}{*}{3.8 (SD=1.48)} & \multirow{2}{*}{5.9 (SD=1.27)} & \multirow{2}{*}{$p=.02^{*}$} \\
    \multirow{2}{*}{Growing Threads} & ``It was easy to find additional papers relevant to each thread using the system.'' & \multirow{2}{*}{4.2 (SD=2.28)} & \multirow{2}{*}{6.1 (SD=0.78)} & \multirow{2}{*}{$p=.04^{*}$} \\
    Organizing References & ``It was easy to organize references into relevant threads using the system.'' & \multirow{2}{*}{3.4 (SD=1.33)} & \multirow{2}{*}{5.6 (SD=1.13)} & \multirow{2}{*}{$p=.007^{**}$} \\
    \bottomrule
    \end{tabular}
    \caption{\new{Descriptions of additional questionnaire items and participants' responses grouped by condition. $p-$values are from two-sided paired samples t-tests. The results suggest that \system helps with collecting clips, growing threads by finding more references, and organizing them efficiently. However, \system did not seem to decrease the demand of the task or of creating and organizing threads. Furthermore, there was no significant difference in terms of technology compatibility/likelihood of adoption between \baseline and \threddy, suggesting a familiarity bias favoring \bs.}}
    \label{table:full_survey}
    \vspace{-2em}
\end{table*}

\subsection*{\new{Appendix C. Vignettes of Participants' Threads}} \label{appendix:thread_vignettes}
\new{The vignettes of threads were simplified by excluding the many clips and references added to each (sub-)thread, and loose yet-to-be organized papers. Note that the structure of threads is subject to change through participants' iteration.}

\noindent\new{\textit{Participant A}'s vignette of threads, simplified.}

\noindent\fbox{\begin{minipage}{.45\textwidth}
\ul{%
  \li{Human-AI collaboration in healthcare}
    \ul{%
       \li{Barriers to AI adoption in healthcare}
       \li{Human AI-onboarding}
       \li{ML as second set of eyes}
       \li{Clinical decision support systems}
    }
   \li{Mental Model for Decision Making and Errors}
   \li{Explainable AI in healthcare}
}
\end{minipage}}
\hfill\\

\noindent\new{\textit{Participant B}'s vignette of threads, simplified.}

\noindent\fbox{\begin{minipage}{.45\textwidth}
\ul{%
  \li{Interpretable model classes and explainability methods}
  \li{Usage of GAMs}
    \ul{%
       \li{GAMs are widely used to detect patterns of data}
    }
   \li{Model interpretability (broadly)}
   \li{Explainable Boosting Machine}
   \li{GAM empirical studies and results}
}
\end{minipage}}
\hfill\\

\noindent\new{\textit{Participant C}'s vignette of threads, simplified.}

\noindent\fbox{\begin{minipage}{.45\textwidth}
\ul{%
  \li{Table-based decision support tools}
    \ul{%
       \li{Sensemaking of collections of online information}
    }
   \li{Review Summarization}
   \li{Aspect Extraction Methods}
   \li{Research on consumer product reviews}
}
\end{minipage}}

\end{document}